\documentclass[%
 reprint,
 amsmath,amssymb,
 aps,
]{revtex4-2}
\usepackage{xr}
\usepackage[colorlinks,linkcolor=red,citecolor=blue,urlcolor=blue]{hyperref}
\usepackage{amsmath}

\usepackage{appendix}
\usepackage{graphicx}% Include figure files
\usepackage{color}
\usepackage{dcolumn}% Align table columns on decimal point
\usepackage{feynmp-auto}
\usepackage{bm}% bold math
\begin{document}

\preprint{APS/123-QED}

\title{Long-range SYK model and boundary SYK model}% Force line breaks with \\

\author{Xiao-Yang Shen}
\affiliation{
Department of Physics, Tsinghua University, Beijing, 100084, China
}%
\affiliation{
Institute for Advanced Study, Tsinghua University, Beijing, 100084, China
}%
\date{\today}% It is always \today, today,

\begin{abstract}
We study a class of long-range solvable models in IR limit which corresponds to a one-dimensional long-range conformal manifold. This class of long-range model can be interpreted as the non-unitary interpolation between the Sachdev-Ye-Kiteav-like model and the free theory. We investigate the chaos and information scrambling of the model by analyzing its out-of-time order correlators. We find the suppression of the Lyapunov exponent by the long-range interaction and a slowdown in butterfly velocity in the emergent light cone which can be interpreted as the contribution from the anomalous dimension of the stress tensor in the spectrum. We further study a Yukawa-SYK model located in a 3-dimensional boundary with a free field living in a 4-dimensional bulk. The boundary IR spectrum of the model contains a tower of the spinning operators protected by the higher spin symmetry of the bulk. We evaluate the central charge and the Lyapunov exponent of the model.
\end{abstract}

\maketitle
\section{Introduction}
%\tableofcontents
Long-range and boundary conformal field theories have been studied for a long time, yet many facts 
stay vague and undetermined. Such “imperfect” theories always turn out to be more realistic in quantum many-body systems like spin models, vector models with long-range interaction or boundary \cite{hiley1965ising,wragg1990ising,cabrera1987role,Lieb2006,dyson1971ising,2021di,2021chai,2023giombi,2022Jchai,2021chakrqborty,2020herzdog,2020Proch,2020Giombi,2023Harribey}, edge modes in the topological phase of matter \cite{1995Callan,2017cho,2017Han,2016WEN,2017Choi}. They also show up in higher energy physics, for instance, spacetime defect in string theory or gravity theory \cite{Recknagel:2013uja,2023Jing,2022Li,2022Geng,2022Geng2,2022Wang,2023Deng}, holographic Kondo model 
\cite{1995affleck,kondo1964resistance,2013Erdmenger} and AdS/BCFT \cite{2011fujita,2022Izumi,2011Takayanagi,Geng:2020qvw,Geng:2021hlu}. The imperfections always raise some intriguing features on such fixed points compared with the conventional CFTs due to the reduction of the symmetry, hence enriching the structure of the theory \cite{2022Schai,2022bianchi,2017herzdog,2023herzdog}. 

To gain insight into such CFTs, people usually apply some analytical perturbative calculations like large N techniques and $\epsilon$ expansion \cite{2016Paulos,2021chai,2020Giombi,2018SLADE}. Recently, modern bootstrap programs are also used to extract some conformal data compatible with the general constraints in the “landscape” of CFTs \cite{2023Levine,2020Behan,2019Behan,2013Liendo,2015Gliozzi}.   

%Apart from the conformal data, the other properties of the long-range and boundary model are also attractive in or away from the IR limit. For example, the chaos and the information scrambling in the chaotic long-range lattice model limits have been investigated in some recent papers, and some exotic emergent light cone structures are discovered due to the power law interaction. One can also look for the transport properties on such model
However, there is still a lack of studies about the boundary and long-range interaction in the strongly coupled CFTs. To investigate those CFTs in a thorough and analytical way, it would be interesting to consider some strongly coupled but solvable long-range and boundary CFTs \cite{2022Popv}. The Sachdev-Ye-Kiteav like models have come into our view as a potential candidate. Sachdev-Ye-Kiteav model is a quantum mechanical model that is solvable under the large N and IR limit \cite{1993PhRvLsachdev,2015Kiteav,2016polchinski,2016PhRvDMaldacena,2017Gross}. The spectrum of the CFT can be solved by considering the four-point function with the ladder structure \cite{2016polchinski,2016PhRvDMaldacena,2017Gross,2017Grossa}. The emergent reparametrization is slightly breaking at low temperatures and the corresponding low energy effective theory is the Schwarzian action, which is the same 
as the boundary dynamical theory in Jakiw-Teitelboim gravity \cite{2018Kiteav,2016Juan,2016Jesen}. The model further displays maximal chaos \cite{2016Maldacena}, which relates to black holes closely. During recent years, many higher dimensional generalization and symmetry-enriched SYK models have been investigated \cite{2020Gu,2021Kim,2017Fu,2017Peng,2017Penga,2018Chang,2017Murugan,chang2023disordered,2018bulycheva,2018Liu,2018DMITRI,2018Khveshchenko,2023Caceres}. The extra dimension and the additional symmetry basically play an important role in diversifying the behaviors of the SYK-like CFTs.
Thus, the SYK-like model with long-range interaction and the model located in the boundary is a good platform for us to dive deep into the strongly coupled long-range and boundary CFTs. The model also sheds some light on the quantum chaos in the presence of long-range interaction \cite{2020Zhou,2023Zhou,2019Chen}.

In this paper, we first study 2d bosonic SYK model with long-range interaction. The IR fixed point corresponds to a conformal manifold parametrized by the power $\alpha$ of the interaction, which interpolates the 2d bosonic SYK (Murugan-Stanford-Witten theory) \cite{2017Murugan} and the free theory. We study the relevant information in the IR spectrum and provide some results about the typical features of the fixed points in the long-range models. We further study the related quantities of the chaos and scrambling, e.g. the Lyapunov exponent and the butterfly velocity by evaluating the out-of-time correlator. Especially, in the light cone limit the butterfly velocity is smaller than the light speed due to the long-range interaction. This indicates that the spatial propagation of the chaos in the long-range system is slower than the conventional unitary CFTs. Furthermore, there exists a critical exponent after which the fast scrambling does not happen and the correlator remains diffusive. We provide the phase diagram of the model.

In the long-range SYK model, the coupling can be regarded as some field living in higher dimensional bulk. Motivated by this thought and the previous work about the holographic Kondo model, we study the Yukawa SYK model in a 3-dimensional boundary with a 4-dimensional bulk free field. The scaling dimension of the field can be tuned by varying the flavor ratio along the IR conformal manifold. When tuning the scaling dimension $\Delta_{\phi} = 1$, there exists a tower of higher spin displacement in the spectrum which is protected by the bulk higher spin symmetry. 
We evaluate the central charge and the hyperbolic chaos exponent of the BCFT.  
\section{long-range SYK model}
Consider the real scalar field $\phi_i$ with flavor index $i = 1,\cdots, N$ in flat $d$ dimensional spacetime. The action is given by
\begin{equation}
    S = \int \mathrm{d}^dx\left[\frac{1}{2}(\partial\phi)^2+J^{i_1,\cdots i_q}\phi_{i_1}\cdots \phi_{i_q}\right]
\end{equation}
\begin{equation}
\langle J_{i_1,\cdots, i_q}\rangle = 0,\quad \langle J_{i_1,\cdots, i_q}^2(x,y)\rangle = (q-1)!\frac{g^2}{N^{q-1}}\frac{1}{|x-y|^{2\alpha}}
\end{equation}
after integrating the random coupling, the action in bilocal form is 
\begin{equation}
    S = \frac{1}{2}\int\mathrm{d}^d x (\partial \phi)^2+\frac{g^2}{N^{q-1}}\int \mathrm{d}^dx\int \mathrm{d}^dy \frac{(\phi^i(x)\phi^i(y))^q}{|x-y|^{2\alpha}}
\end{equation}
the first term is a local kinetic term, giving the field free scaling dimension $[\phi] = \frac{d-2}{2}$. The second term is a bi-local $O(N)$ interaction with the long-range interaction strength decay with $|x-y|^{-2\alpha}$. The scaling dimension of $[g^2] = 2 d-q(d-2)-2 \alpha$. 
When $\alpha \to 0$, from the perspective of the bi-local field, the interaction is mostly non-local, involving two fields interacting at different distances with the same weight. When $\alpha\to \infty$, the bi-local term becomes local, which is similar to the Brownian SYK model \cite{2018Saad}. To make the second term relevant in IR, we restrict ourselves to $0\leq\alpha<d-q(d-2)/2$.

\begin{figure}[h]
    \centering
    \includegraphics[width = 0.48\textwidth]{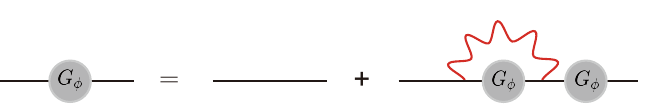}    
    \caption{Schwinger-Dyson equation in large N limit when $q = 2$. $G_\phi$ is the dressed propagator. Red wavy line is the variance of disordered coupling. The equation is diagrammatically similar to the massless
fundamental fermions coupled to Chern-Simons gauge field in the light cone gauge and large N limit \cite{2012Giombi,2012Aharony}.}
\label{2ptp}
\end{figure}

To solve the model, we first take the large-N limit and then focus on the low energy limit. 
The Schwinger-Dyson equation is plotted in Fig.(\ref{2ptp}).
Under large N limit, the Dyson-Schwinger equation reads:
\begin{equation}
    G^{-1}(p)=p^2-\Sigma(-p)
\end{equation}
and the self energy is given by:
\begin{equation}
    \Sigma(x, y)=g^2 \frac{G(x, y)^{q-1}}{|x-y|^{2 \alpha}}
\end{equation}
In the low-energy limit, we ignore the free propagator and pick an ansatz $G(x) = \frac{b}{|x|^{2\Delta
_{\phi}}}$.
The solution of the equation then goes to:
\begin{equation}\label{2pt}
    \Delta_\phi=\frac{d-\alpha}{q}, \quad g^2 b^q=-c_d\left(\frac{d-\alpha}{q}\right)^{-1} c_d\left(\frac{q-1}{q} d+\frac{\alpha}{q}\right)^{-1}
\end{equation}
here we denote
\begin{equation}\label{Fourier_coe}
c_d(a) \equiv \frac{\pi^{d / 2}}{2^{2 a-d}} \frac{\Gamma\left(\frac{d}{2}-a\right)}{\Gamma(a)}
\end{equation}

\section{spectrum}
We focus on the average four-point function in large-N limit
\begin{equation}
\label{vaccumm_4pt}
\begin{aligned}
 &\frac{1}{N^2}\left\langle\phi^i(x_1) \phi^i(x_2) \phi^j(x_3) \phi^j(x_4)\right\rangle\\
  &\equiv G(x_{12})G(x_{34})+\frac{1}{N}\mathcal{F}(x_1, x_2, x_3, x_4) 
  \end{aligned}
\end{equation}
In the large-N limit, the four-point function processes a similar ladder structure as the SYK model
\begin{equation}
    \mathcal{F} = \sum_{n = 0}^{\infty}K^{\star n}\star F_0 = \frac{F_0}{1-K}
\end{equation}
$F_0$ denotes the zero-rung kernel. The star $\star$ symbolizes the integration of the mediate coordinates
\begin{equation}
    K\star F \equiv \int \mathrm{d}^d x_{2}\int \mathrm{d}^d x_{3} K(x_1,x_2,x_3,x_4)F(x_1,x_2,x_3,x_4)
\end{equation}
Hence, the information about the spectrum conceals in the recursive kernel $K$. We are about to read the spectrum by evaluating the eigenvalue of the kernel
\begin{equation}
K=(q-1) g^2 G\left(x_{12}\right) \frac{G(x_{23})^{q-2}}{\left|x_2-x_3\right|^{2 \alpha}} G\left(x_{34}\right)
\end{equation}
In the low-energy limit, the conformal propagator $G$ is given by Eq.(\ref{2pt}). The eigenvalue of the kernel can be evaluated by applying the eigenvector $\mathcal{V}$
\begin{equation}
    \mathcal{V}\star K = k(\Delta
    ,\ell) K
\end{equation}
\begin{equation}
    \mathcal{V} = |x|^{\Delta-\ell-2 \Delta_\phi}\left(x_{\mu_1} \ldots x_{\mu_\ell}-\text{traces}\right)
\end{equation}
For now, we focus on $d = 2$. The permitted range of $\alpha$ is now $0<\alpha<2$. The eigenvalue is given in Eq.(\ref{spectrum_blvm})
\begin{widetext}
\begin{equation}\label{spectrum_blvm}
k(\Delta, \ell)=(q-1) \frac{(-2+\alpha+q)^2}{q^2} \frac{\Gamma\left(1-\frac{2-\alpha}{q}\right)^2 \Gamma\left(\frac{2-\alpha}{q}-\frac{\Delta-\ell}{2}\right) \Gamma\left(\frac{\Delta+\ell}{2}+\frac{2-\alpha}{q}-1\right)}{\Gamma\left(\frac{2-\alpha}{q}\right)^2 \Gamma\left(2-\frac{2-\alpha}{q}-\frac{\Delta-\ell}{2}\right) \Gamma\left(\frac{\Delta+\ell}{2}-\frac{2-\alpha}{q}+1\right)}
\end{equation}
\begin{figure}[t]
    \centering
\includegraphics[width = 1.
\textwidth]{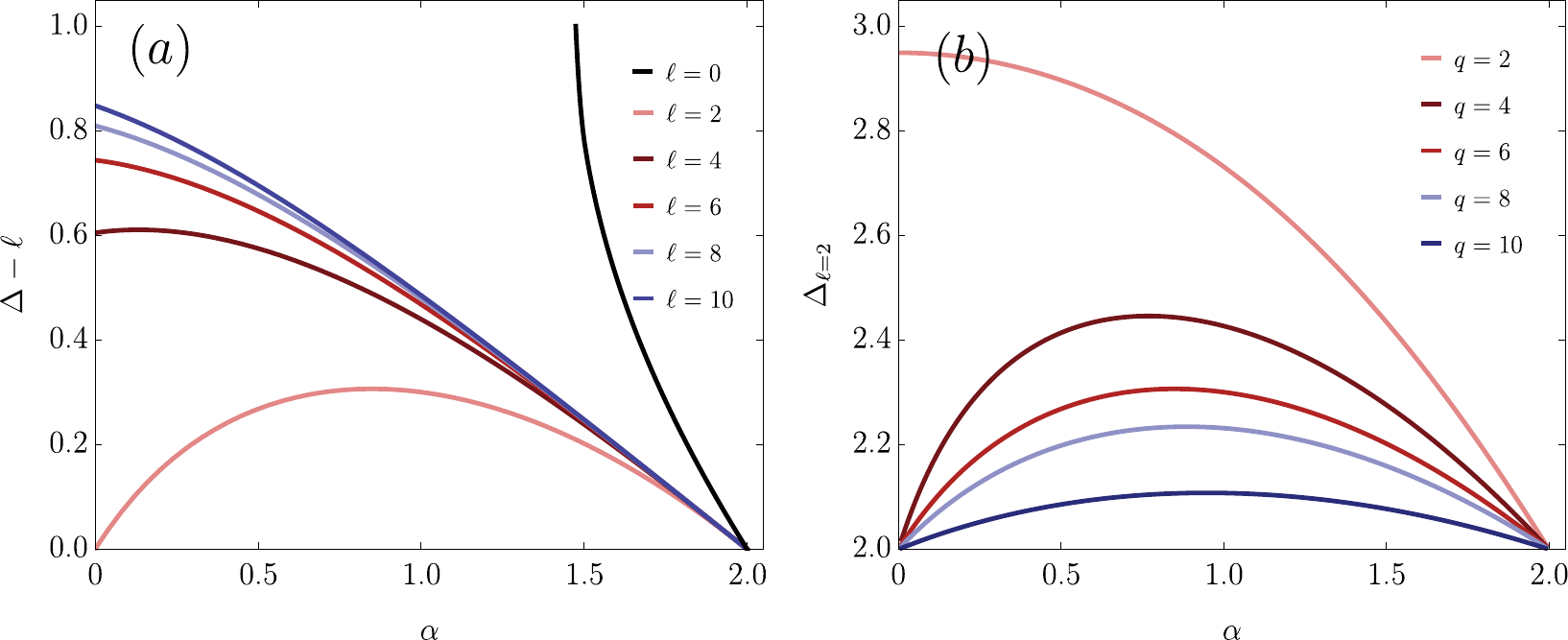}
    \caption{ Operator spectrum for 2d long-range bosonic model.  (a) is the leading twist for operators of $\ell = 0,2,4,6,8,10$ when $q = 4$. The $l = 0$ spectrum is complex for $\alpha<\alpha_c = 1.475$. For $\ell = 2$, the operator acquires a non-zero twist when $0<\alpha<2$ and the twist vanishes at $\alpha = 0,2$ (MSW and free theory) indicating the CFT is unitary. The spectrum is bounded by $\Delta-\ell = 1-\alpha/2$ when $\ell\to \infty$. (b) is the leading twist for $\ell = 2$ as function of $\alpha$ and $q$. The twist gets smaller when $q$ is larger. When $q = 2$, the MSW model becomes a trivial mass term by field redefinition, hence we expect a jump in the dimension when $\alpha$ becomes non-zero.}
    \label{spectrum_blvm_1}
\end{figure}
\end{widetext}
The spectrum can be derived by solving the equation:
\begin{equation}
    k(\Delta,\ell) = 1
\end{equation}

We focus on the operator with the leading twist, as in Fig.(\ref{spectrum_blvm_1}).  When $\alpha < 1.4741$, the spin-0 channel is complex, which is also observed in the MSW model and 3d bosonic disordered theory \cite{2017Murugan,CC2022}
, this indicates the potential is not stable and bounded from below. When $\alpha$ exceeds the $\alpha_c$, the dimension becomes real and decreases from 1 to 0. For $\ell = 2$, we find that the spectrum with a leading twist acquires an anomalous dimension, which means that the stress tensor does not exist. This symbolizes the breakdown of local conservation due to the presence of long-range interaction and is a general signature in the long-range fixed point \cite{2021chai,2017PhRBEHAN}.  Another way to interpret this is that in the model the coupling $J(x)$ is dealt with at the meanfield level, if $J(x)$ is taken to be some physical field (with fixed scaling dimension), one can again evaluate the spectrum and find the stress tensor of the model does not receive an anomalous dimension.

To fix the problem of the complex $\ell = 0$ spectrum, we can also enhance the model to the $\mathcal{N}(2,2)$ version. The superpotential of the model is given by
\begin{equation}
\mathcal{W}=\int d^2 \theta \int d^2 x J^{i_1,\cdots i_q}\Phi_{i_1}\cdots \Phi_{i_q}+\text{c.c}
\end{equation}
The variance of the random variable is
\begin{equation}
    \langle J^2_{i_1,\cdots ,i_q}(X,Y)\rangle \propto (q-1)!\frac{g^2}{N^{q-1}}\frac{1}{|X-Y|^{2\alpha}}
\end{equation}
$X,Y$ are the supercoordinates in $\mathcal{N}(2,2)$ convention. For the detailed convention of superspace, see \cite{chang2023disordered}.
The conformal dimension of the chiral superfield $\Delta_{\Phi} = \frac{1-\alpha}{q}$. 
The ladder kernel function $k(\Delta,\ell)$ is 
\begin{equation}
(q-1) \frac{\Gamma\left(1-\Delta_{\Phi}\right)^2}{\Gamma\left(\Delta_{\Phi}\right)^2} \frac{\Gamma\left(\frac{\Delta+\ell}{2}+\Delta_{\Phi}\right)}{\Gamma\left(1+\frac{\Delta+\ell}{2}-\Delta_{\Phi}\right)} \frac{\Gamma\left(\frac{-\Delta+\ell}{2}+\Delta_{\Phi}\right)}{\Gamma\left(1+\frac{\ell-\Delta}{2}-\Delta_{\Phi}\right)}
\end{equation}

\begin{figure}[h]
    \centering
\includegraphics[width = 0.495\textwidth]{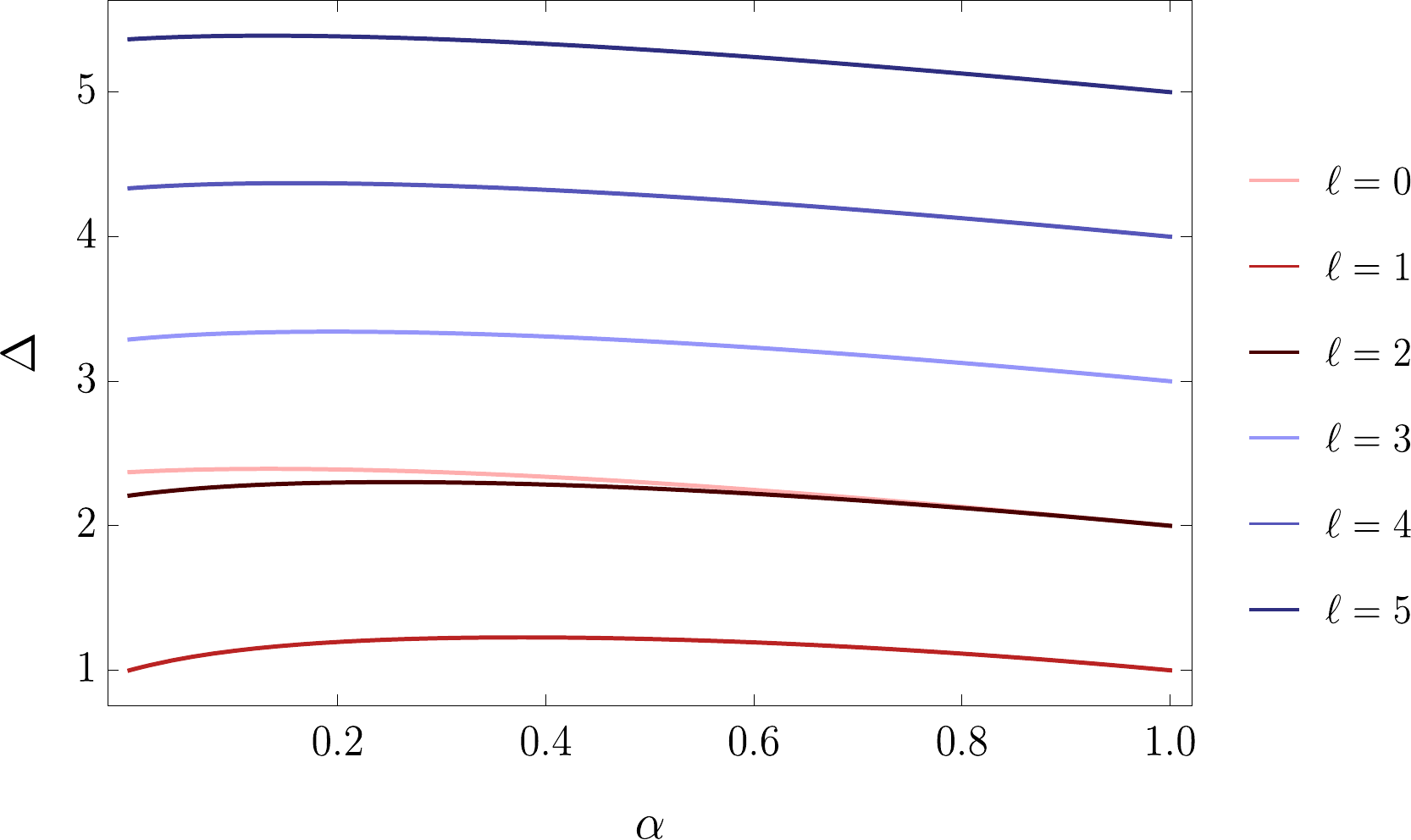}
    \caption{The leading twist in the $\mathcal{N} = (2,2)$ SUSY operator spectrum with spin $0,1,2,3,4,5$. When $\alpha = 0,2$ the $\ell = 1$ curve stands for the $\mathcal{N} = 2$ multiplet of superconformal current, which contains the supercurrent, stress tensor, and R charge.}
    \label{fig:spectrum_susy}
\end{figure}

The operator spectrum for SUSY theory is given in Fig.(\ref{fig:spectrum_susy}).

\section{Chaos and scrambling}
The normalized out-of-time order correlator is defined to be:
\begin{equation}
\frac{\left\langle\phi_i\left(\frac{3 \beta}{4}+i t, x\right) \phi_j\left(\frac{\beta}{2}, 0\right) \phi_i\left(\frac{\beta}{4}+i t, x\right) \phi_j(0,0)\right\rangle}{\left\langle\phi_i(\frac{3\beta}{4},0)\phi_i(\frac{\beta}{4},0)\right\rangle\left\langle\phi_j(\frac{\beta}{2},0)\phi_j(0,0)\right\rangle}
\end{equation}
here the $\langle\cdot\rangle$ denotes the thermal expectation. A complication here is how we define temperature in this system since the energy is not conserved in the long-range model. A typical model is the Brownian SYK with infinite-range interactions, in which we consider the infinite temperature correlation function \cite{2018Saad}. A possible but less strict thought is to consider $J$ as a “classical” field instead of the coupling and define temperature in the new theory. Let us first review a direct way to study the general form of the OTO correlator under this set up. In 2d large-N thermal CFT, we are able to map this correlator to a zero-temperature vacuum correlator Eq.(\ref{vaccumm_4pt}) via the conformal mapping
\begin{equation}
    z=e^{\frac{2 \pi}{\beta}(x+i \tau)}
\end{equation}
starting from the Euclidean four point function, we analytically continue it to the real time, meanwhile keeping track of the cross ratio $\chi$ and $\bar{\chi}$. During the continuation, $\chi$ crosses the branch cut and enters the second sheet, hence triggering the multi-valueness of the hypergeometric function in the OPE expansion. At large t, the cross ratio in the vacuum correlator is
\begin{equation}
    \chi \rightarrow-4 i e^{-t+x}, \quad \bar{\chi} \rightarrow-4 i e^{-t-x}
\end{equation}
We are interested in the behavior of the conformal block when the $t$ is large, thus we cast the expression into the conformal block 
\begin{equation}
    G_{\Delta,\ell} \propto \chi^{1-\frac{\Delta+\ell}{2}} \bar{\chi}^{\frac{\Delta
    -\ell}{2}} \propto e^{-\frac{\pi}{2}\ell i}e^{t(\ell-1)-x(\Delta-1)}
\end{equation}
this basically tells us that the operators with $\text{Re}(\ell)>1$ in the spectrum is responsible for the $e^{\lambda t}$ in the OTO correlator \cite{2015Roberts}.

When studying the conformal four-point function, we find a basis of eigenvectors  $\Phi_{\Delta,\ell}$ labeled by dimensions and spin which are also the eigenvectors of the Casimir operator. The so-called conformal partial waves $\Phi_{\Delta,\ell}$ is a linear superposition of the conformal blocks and the shadow block in 1+1d CFT
\begin{equation}
    \Phi_{\Delta
    ,\ell}\equiv S_{\Delta
    ,\ell}G_{2-\Delta,\ell}+S_{2-\Delta
    ,\ell}G_{\Delta,\ell}
\end{equation}

Under the expansion, the conformal four-point function is given by:
\begin{equation}
\begin{aligned}
\label{expasion}
\mathcal{F}& =\sum_{\ell \text { even }}^{\infty} \int_0^{\infty} \mathrm{d} s \frac{\left\langle\Phi_{\Delta, \ell}, \mathcal{F}_0\right\rangle}{1-k(\Delta, \ell)} \frac{\Phi_{\Delta, \ell}}{\left\langle\Phi_{\Delta, \ell}, \Phi_{\Delta, \ell}\right\rangle}\\
& =-\frac{\pi}{2}\int_{-\infty}^{\infty} \mathrm{d} s  \int_{\mathcal{C}} \frac{d \ell}{2 \pi i} \frac{1+e^{-i \pi \ell}}{\sin (\pi \ell)}\frac{\left\langle\Phi_{\Delta, \ell}, \mathcal{F}_0\right\rangle}{1-k(\Delta, \ell)} \frac{S_{2-\Delta
,\ell}G_{\Delta, \ell}}{\left\langle\Phi_{\Delta, \ell}, \Phi_{\Delta, \ell}\right\rangle}
\end{aligned}
\end{equation}
here we have $\Delta
 =1+is$ and unfold the integral of $s$ by using the shadow symmetry in conformal partial waves and the contour integration of $\ell$ circles around the even spin pole. We carefully deform the contour of $\ell$ to the imaginary direction with $\text{Re}(\ell)<1$ and pole with $\text{Re}(\ell)>1$ encounter during the deformation. Such poles can be proven to be totally
come from the $\frac{1}{1-k}$ and the other parts in Eq.(\ref{expasion}) do not contribute. In general, the other part is model-independent and can be calculated in the generalized free field model. 
Substituting $s\to -p$, the contribution of a pole is given by:
\begin{equation}\label{oto}
    \mathcal{F}\supset-\frac{\pi}{2}\int \mathrm{d}p \frac{e^{t(\ell-1)+ipx}}{\sin\frac{\pi\ell}{2}}\left.\frac{1}{\partial_\ell k(\Delta
    ,\ell)}\right|_{k = 1} \frac{\left\langle\Phi_{\Delta, \ell}, \mathcal{F}_0\right\rangle S_{2-\Delta
    ,\ell}}{\left\langle\Phi_{\Delta, \ell}, \Phi_{\Delta, \ell}\right\rangle} 
\end{equation}
Notice that here $\ell(p)$ is the function of $p$ and is decided by the condition $k = 1$. For the SUSY case, one can do the same thing by using the superconformal partial waves $\Xi_{\Delta
,\ell}$ \cite{chang2023disordered}. 
\begin{figure}[h]
    \centering
\includegraphics[width = 0.5\textwidth]{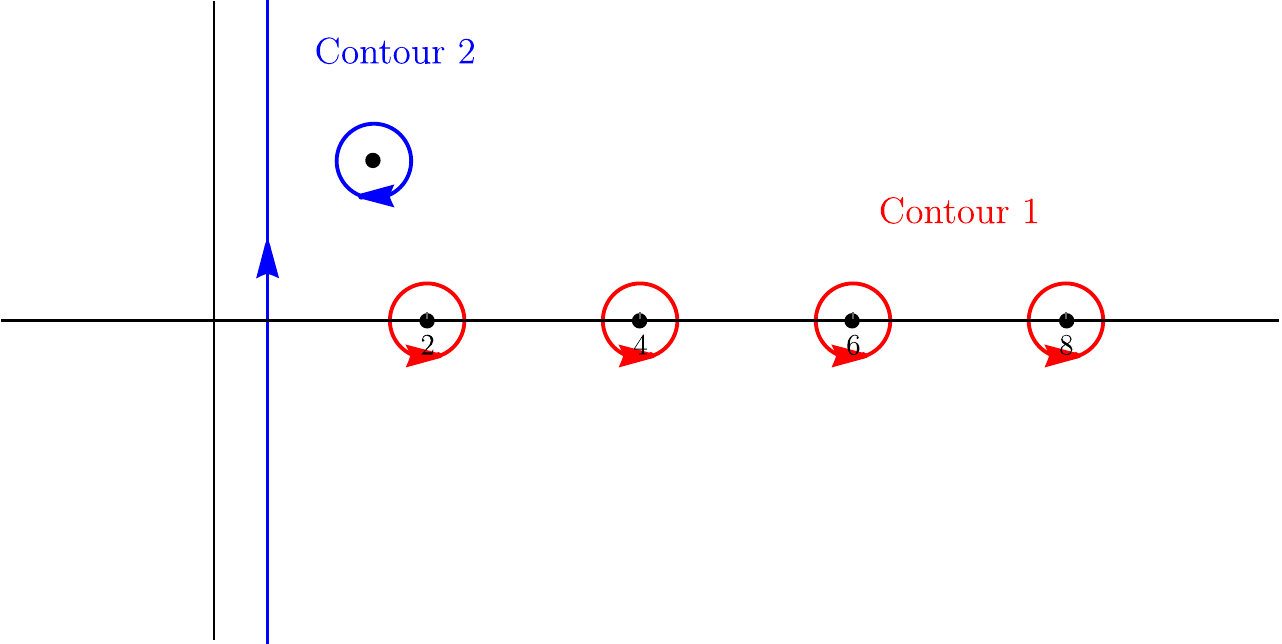}
    \caption{Sommerfield-Watson resummation-like procedure \cite{2012Costa}. One deforms contour 1 smoothly to contour 2 with isolated poles given in $k(\ell,s) = 1$ and a line integration contour with $\text{Re}(\ell)<1$ along the imaginary axis in the complex $\ell$ plane. In our case, we find a pole located in the real axis.}
    \label{fig:enter-label}
\end{figure}

\begin{figure}[h]
    \centering
\includegraphics[width = 0.48\textwidth]{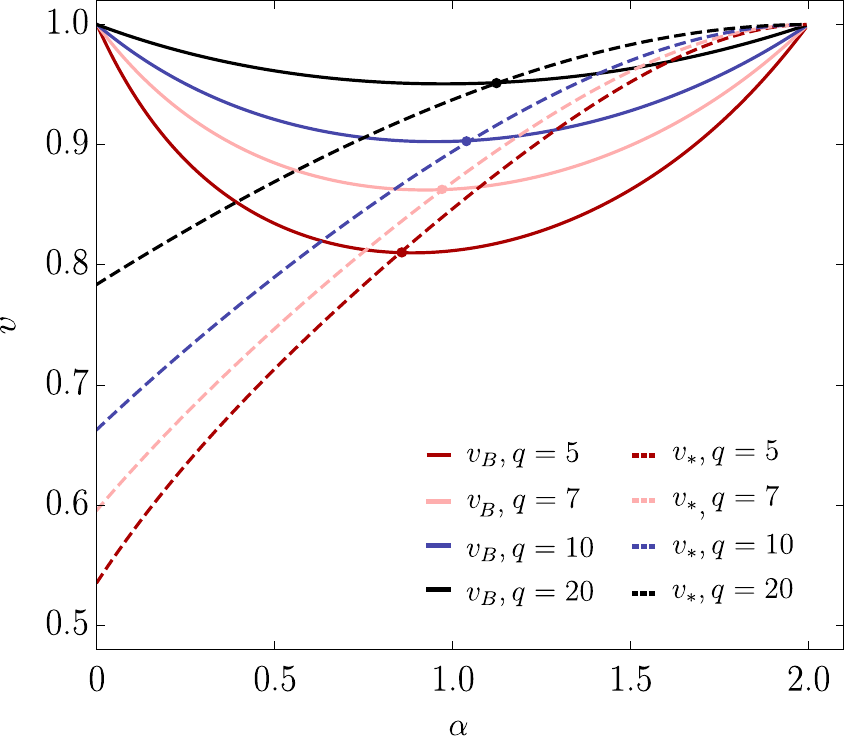}
    \caption{$v_*$ and $v_B$ as function of $\alpha$ for some values of $q$.}
    \label{fig:velocity}
\end{figure}

Starting from the expression of the Eq.(\ref{oto}), we discuss two different cases when carrying the $p$ integration, which corresponds to the Regge limit and light cone limit in CFT \cite{2017Murugan,2019Gu,2023Choi,2020Mark}.
\begin{enumerate}
    \item $\boldsymbol{x}\ll \boldsymbol{t}$. In this case, the integral can be regarded as the Gaussian integral central in $p = 0$. The increasing part processes a diffusion profile:
    \begin{equation}
       \frac{1}{N}\mathcal{F}\sim -\frac{1}{N\sqrt{t}}\exp \left((\ell(0)-1) t-\frac{x^2}{2 t \ell^{\prime \prime}(0)}\right)
    \end{equation}
    \begin{widetext}

    \begin{figure}[t]
    \centering
\includegraphics[width = 1\textwidth]{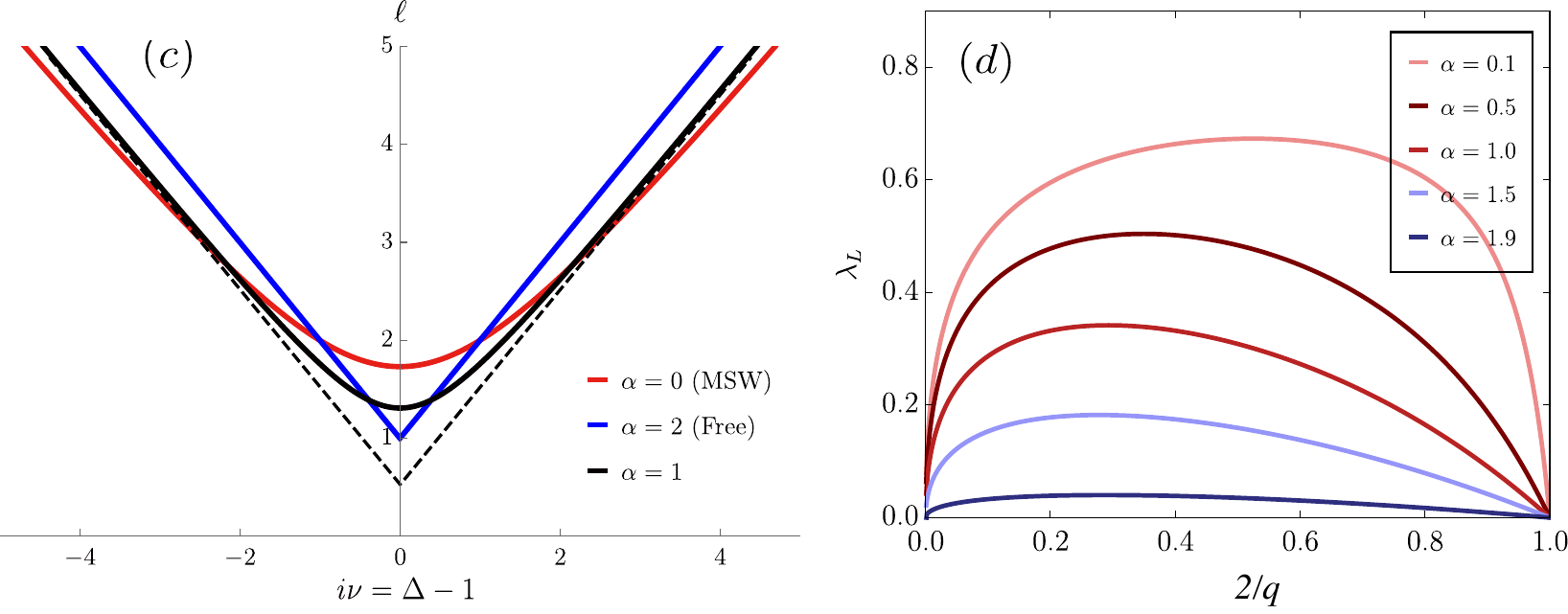}
    \caption{Fig.(c) is the Regge trajectory formed by the leading twist operator in the principle series. The blue line and the red line are the free theory and SYK (MSW) model. When $0<\alpha<2$, the trajectory circumvents the pole of stress tensor $(1,2)$, hence the spin-2 operator acquires a twist. The black dashed line is the behavior when $\ell\to \infty$, in which the anomalous dimension of the operators vanishes. The Lyapunov exponent is defined by the Regge intercept minus 1, which interpolates between the free theory and the SYK case. Fig.(d) is the Lyapunov exponent as functions of $q$ and $\alpha$. From the point of local action, the long-range interaction hampers the chaotic behavior.}
    \label{fig:chaos}
\end{figure}

\end{widetext}
    $\ell(0)-1$ is defined to be the Lyapunov exponent. From the point of CFT, $\ell(0)$ is the Regge intercept in Fig.(\ref{fig:chaos}) and can be extracted from:
    \begin{equation}
        k(1,\ell(0)) = 1
    \end{equation}
    \item $\boldsymbol{x}\sim\boldsymbol{t}$. In this case, the pole in the factor $\sin(\frac{\pi\ell}{2})$ dominates the integral, and the integral can be evaluated by taking the residue of the pole:
\begin{equation}
     \frac{1}{N}\mathcal{F}\sim -\frac{1}{N}\exp\left(t-x/v_B\right)
\end{equation}
   $v_B\equiv i/p_*$ is dubbed the butterfly velocity, $p_*$ is decided by $k(1-ip_*,2) = 1$.  
\end{enumerate}
When we increase the $v = x/t$, it is possible to find a transition between these two behaviors. Whether the transition happens depends on a critical velocity $v_*$ where the pole contribution exceeds the saddle point contribution:
\begin{equation}
v_* \equiv i\partial_p \ell(ip)|_{p^*} = -\left. i\partial_{p}k(1-ip,\ell)/\partial_\ell k(1-ip,\ell)\right|_{p=p_*,\ell = 2}
\end{equation}
When $v_*<v_B$, the region dominated by the pole always exists thus the fast scrambling exists whereas for $v_*>v_B$ the pole does not dominate and the fast scrambling is absent. We plot the $v_*$ and $v_B$ for different $q$ and $\alpha$, as shown in Fig.(\ref{fig:velocity}).
We find the $v_B$ is smaller than 1 and only equals 1 in SYK and free limit. In the conventional unitary 1+1 dimension CFT, the leading $\ell = 2$ operator is always the stress tensor with dimension $\Delta = 2$ saturating the unitarity bound. Hence the butterfly velocity is always equal to the light speed in 1+1d Lorentz invariant critical point \cite{2021Kim}. However in the presence of the long-range interaction, due to the breakdown of the local conservation, the stress tensor acquires a non-trivial anomalous dimension which shifts the location of the pole, hence resulting in a slower butterfly velocity in the emergent light cone.

\begin{figure}[h]
    \centering
\includegraphics[width = 0.48\textwidth]{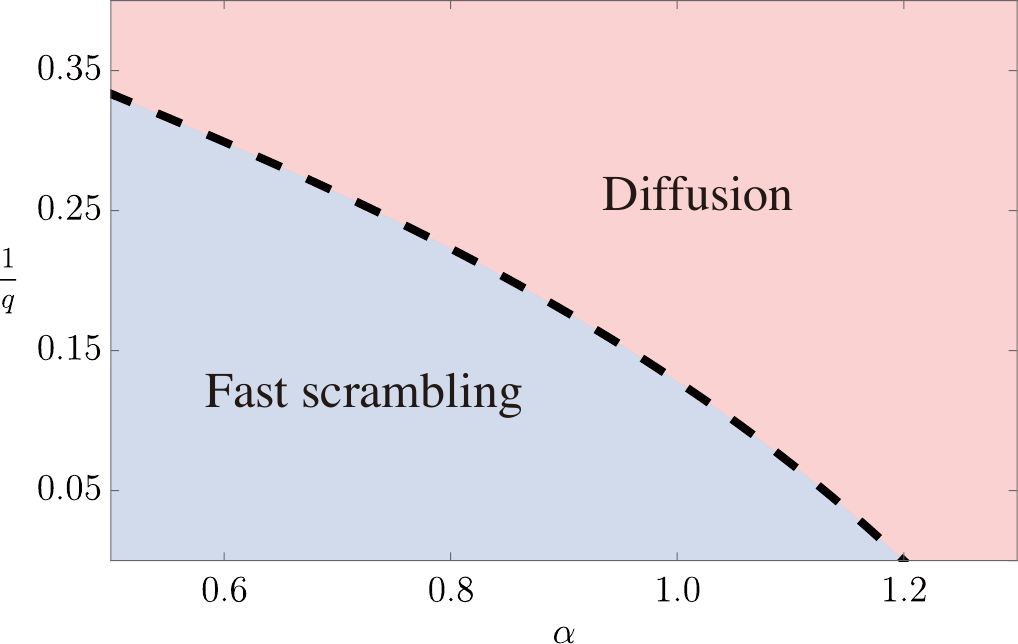}
    \caption{Phase diagram of the long-range SYK model as functions of $\alpha$ and $q$. Extrapolation is used for general value of $q$. For $q = 3$ and $q \to \infty$, $\alpha_c = 0.5$ and $1.2$. }
    \label{fig:pd}
\end{figure}

The behavior of $v_*$ has several intriguing features. when $\alpha \to 0$, $v_*$ equals to explicit bosonic MSW result \cite{2020Mark}:
\begin{equation}
v_* = \frac{(-1+\Delta_{\phi})^2}{-1+(-2+\Delta_{\phi})[(-1+\Delta_{\phi}) \pi\Delta_{\phi} \cot(\pi \Delta_{\phi})-2\Delta_{\phi}]}
\end{equation}
$v_*$ keeps increasing with $\alpha$ and becomes $1$ in the free limit. In small $\alpha$, the $v_B$ is always greater than $v_*$, hence guarantee a region $t\in \left(\frac{x}{v_B},\frac{x}{v_*}\right)$ where the theory is fast scrambling. There exists a transition point $0<\alpha_c<2$ for all range of $q$, after which the fast scrambling does not happen, see Fig.(\ref{fig:pd}).

\section{boundary SYK model}

In the above context, we discuss an SYK-like long-range model, in which the disordered coupling can be viewed as some sort of higher dimensional field originating from the higher dimensional conformal field theory.
This feature is quite similar to the holographic Kondo effect. In the disordered context, we can also try to construct some theories, with SYK-like interaction located in the boundary. A non-trivial example in UV is given as follows:
\begin{equation}\label{bsyk}
    S = \int d^dx \left(\frac{1}{2}(\partial\phi)^2+g^a_{ij}\sigma^a\phi_i\phi_j\right)+\int d^{d+1} x\frac{1}{2}(\partial \sigma)^2
\end{equation}
This is a Yukawa-SYK theory living the boundary of the $d+1$ spacetime and $\sigma$ lives in the bulk. The $g^a_{ij}$ is the random coupling with variances $g^2$, the flavors of $\phi$ and $\sigma$ are $M$ and $N$ and we fix the ratio $\lambda = \frac{M}{N}$ when assuming the large $N$ limit. The bulk free propagator in the presence of the boundary can be derived via the image method:
\begin{equation}
\begin{aligned}
    G^0_{\phi} = \frac{\Gamma\left(\frac{d+1}{2}\right)}{2(d-1) \pi^{\frac{d+1}{2}}}\left(\frac{1}{\sqrt{(x-x^\prime)^2+(y-y^\prime)^2}^{d-1}}\right.\\
    \left.+\frac{\gamma}{\sqrt{(x-x^\prime)^2+(y+y^\prime)^2}^{d-1}}\right)
\end{aligned}
\end{equation}
Here $x$ is the $d$ dimensional coordinates parallel to the bulk and $y$ is the coordinates perpendicular to the bulk. The Neumann boundary condition $\gamma = 1$ is used here. We apply the Fourier transformation along the $x$ direction and the propagator in momentum space reads
\begin{equation}
    G^0_{\phi}(p)=\frac{e^{-p|y-y^{\prime}|}+e^{-p(y+y^{\prime})}}{2 |p|}
\end{equation}
Thus in the boundary limit $y,y^\prime\to 0$, the free propagator can be treated as $\frac{1}{|p|}$. The self-energy of the d-dimensional Yukawa-SYK model is given as:
\begin{equation}
    \Sigma_{\sigma} = \frac{1}{2}g^2G_{\phi}^2(x),\quad \Sigma_{\phi} = g^2\lambda G_{\phi}(x)G_{\sigma}(x)
\end{equation}
and the Schwinger-Dyson equation reads:
\begin{equation}
    G_\phi = \frac{1}{p^2-\Sigma_{\phi}},\quad G_{\sigma} = \frac{1}{|p|-\Sigma_{\sigma}}
\end{equation}
$\frac{1}{p}$ here is the free $\sigma$ propagator when converting to the boundary. In the IR limit, we first ignore the free part and derive the IR solution:
\begin{equation}
2\Delta_{\phi}+\Delta_{\sigma} = d,
\end{equation}
\begin{equation}
    \lambda = \frac{c_d(2\Delta_{\phi})c_d(d-2\Delta_{\phi})}{2c_d(\Delta
    _{\phi})c_d(d-\Delta_{\phi})}
\end{equation}
\begin{equation}
g^2b_{\phi}^2b_{\sigma} = -\frac{2}{c_d(2\Delta_{\phi})c_{d}(d-2\Delta_{\phi}))}
\end{equation}
$c_d(a)$ is given in Eq.(\ref{Fourier_coe}). The permitted range for $\Delta_{\phi}$ to ensure the interaction dominate in IR is: 
\begin{equation}
    \frac{d-2}{2}\leq \Delta_{\phi} \leq \frac{d+1}{4}
\end{equation}

The four-point function is given by the summation ladder diagram, and the spectrum is again given by 
 the kernel:
\begin{equation}\label{kernel}
    K = \left(\begin{array}{cc}
\lambda g^2 b_\phi^2 b_\sigma k_\phi & \lambda g^2 b_\phi^3 k_\phi \\
g^2 b_\sigma^2 b_\phi k_\sigma & 0
\end{array}\right)
\end{equation}
where 
\begin{equation}
    k_{i} = \frac{(-1)^\ell+1}{2}\frac{\pi^d\Gamma(\frac{d}{2}-\Delta_{\phi})^2\Gamma(\Delta_i-\frac{\Delta-\ell}{2})\Gamma(\frac{\Delta+\ell}{2}+\Delta_i-\frac{d}{2})}{\Gamma(\Delta_{\phi})^2\Gamma(d-\frac{\Delta-\ell}{2}-\Delta_{i})\Gamma(\frac{\Delta+\ell}{2}-\Delta_i+\frac{d}{2})}
\end{equation}
where $k_\phi,k_\sigma$ denotes the eigenvalue of the kernel with the entries to be $G_{\phi},G_{\sigma}$, respectively. The ladder kernel function can be derived from the eigenvalue of kernel Eq.(\ref{kernel})
\begin{equation}\label{ladder_kernel_function}
    k(\Delta,\ell) = g^2\lambda b_\phi^2b_{\sigma} k_\phi(1+g^2b_{\phi}^2b_{\sigma}k_\sigma)
\end{equation}

We then obtain the spectrum of the model by solving $k(\Delta,\ell) = 1$.
\begin{figure}[h]
    \centering
\includegraphics[width = 0.48\textwidth]{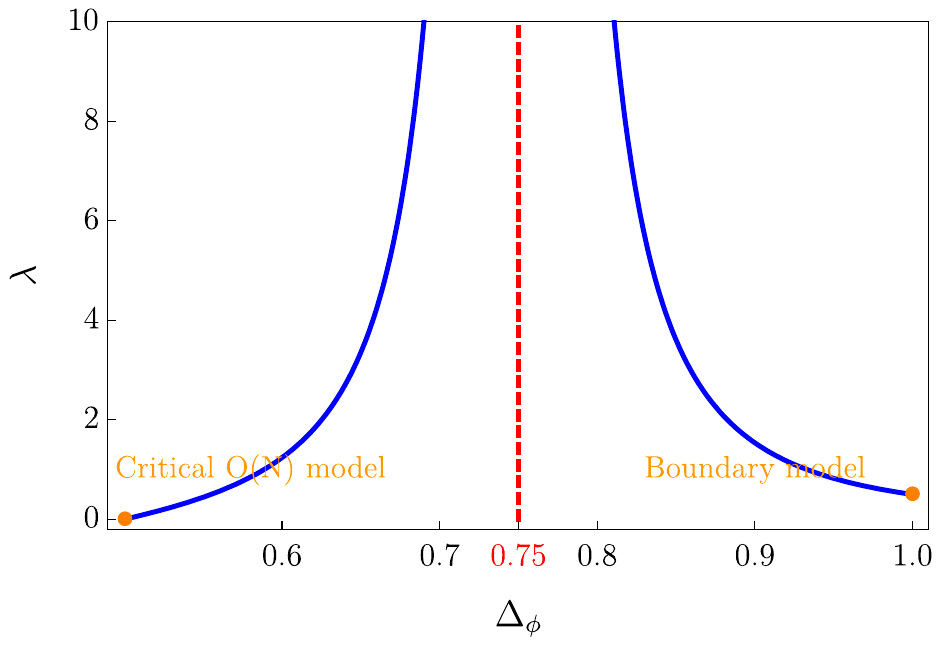}    \caption{$\lambda$ as the function of $\Delta_{\phi}$. There are two branches of $\lambda-\Delta_{\phi}$ separated by $\Delta_{\phi} = \frac{3}{4},\Delta_{\sigma} = \frac{3}{2}$. When $\Delta_{\phi} = \frac{1}{2}$, the model is critical $O(N)$ vector model by field redefinition. We find $\Delta_{\phi} = 1$ sitting in another branch of conformal manifold and corresponds to $\lambda = \frac{1}{2}$. At this point, we find a tower of higher spin displacement operator.}
    \label{fig:lambda_delta_curve}
\end{figure}
We will focus on the physical dimension $d = 3$ in the following context. In this case, 

\begin{equation}
    \lambda = \frac{\left(\Delta_\phi-2\right)\left(2 \Delta_\phi-3\right)\left(1+\sec \left(2 \pi \Delta_\phi\right)\right)}{4\left(2 \Delta_\phi\left(4 \Delta_\phi-5\right)+3\right)}
\end{equation}
\begin{equation}
    g^2 b_{\phi}^2b_{\sigma} = \frac{16}{\pi^3}\left(\Delta_\phi-1\right)\left(\Delta_\phi-\frac{1}{2}\right)\left(\Delta_\phi-\frac{3}{4}\right) \cot \left(2 \pi \Delta_\phi\right)
\end{equation}

$\lambda$ is plot as a function with scaling dimension $\Delta_{\phi}$, see Fig.(\ref{fig:lambda_delta_curve}).

Notice that when $\Delta_{\phi} = \frac{1}{2}$, $\lambda = 0$, the model corresponds to the critical $O(N)$ model \cite{CC2022}. The solution splits into two branches, the first branch connects with the critical $O(N)$ vector model, and is divergent when $\Delta_{\phi} = \frac{3}{4}$ when the mass term of $\sigma$ becomes marginal. The other branch lives with $\Delta_{\phi}>\frac{3}{4}$, and  $\Delta_{\sigma}<\frac{3}{2}$, hence can be regarded as some free $\sigma$ field living in the higher dimensional space-time.  When $\Delta_{\phi} = 1, \Delta_{\sigma} = 1$, we have $\lambda = \frac{1}{2}$. However, the solution is different from the $q = 3$ bosonic SYK model, even though $\Delta_{\sigma} = \Delta_{\phi} = 1$. In fact, the spectrum ladder function has the following relation with $q = 3$ bSYK model \cite{2018Liu}. 
\begin{equation}\label{ladder_kernel_function}
    k(\Delta,\ell) = 1-(1-k_{q = 3\text{bSYK}})(1+\frac{1}{2}k_{q = 3\text{bSYK}})
\end{equation}
 $k_{q = 3\text{bSYK}}$ is the ladder kernel function for $q = 3$ bSYK model:
\begin{equation}\label{ladder_kernel_bsyk}
    k_{q = 3\text{bSYK}} = \frac{1}{\pi^4}k_{\phi}(\Delta_{\phi} = 1)
\end{equation}
$\frac{1}{\pi^4}$ comes from the combination of coefficients $J^2b_{\phi}^3$. The following expression means that the spectrum of bosonic SYK is included.

\begin{figure}[h]
    \centering
\includegraphics[width = 0.49\textwidth]{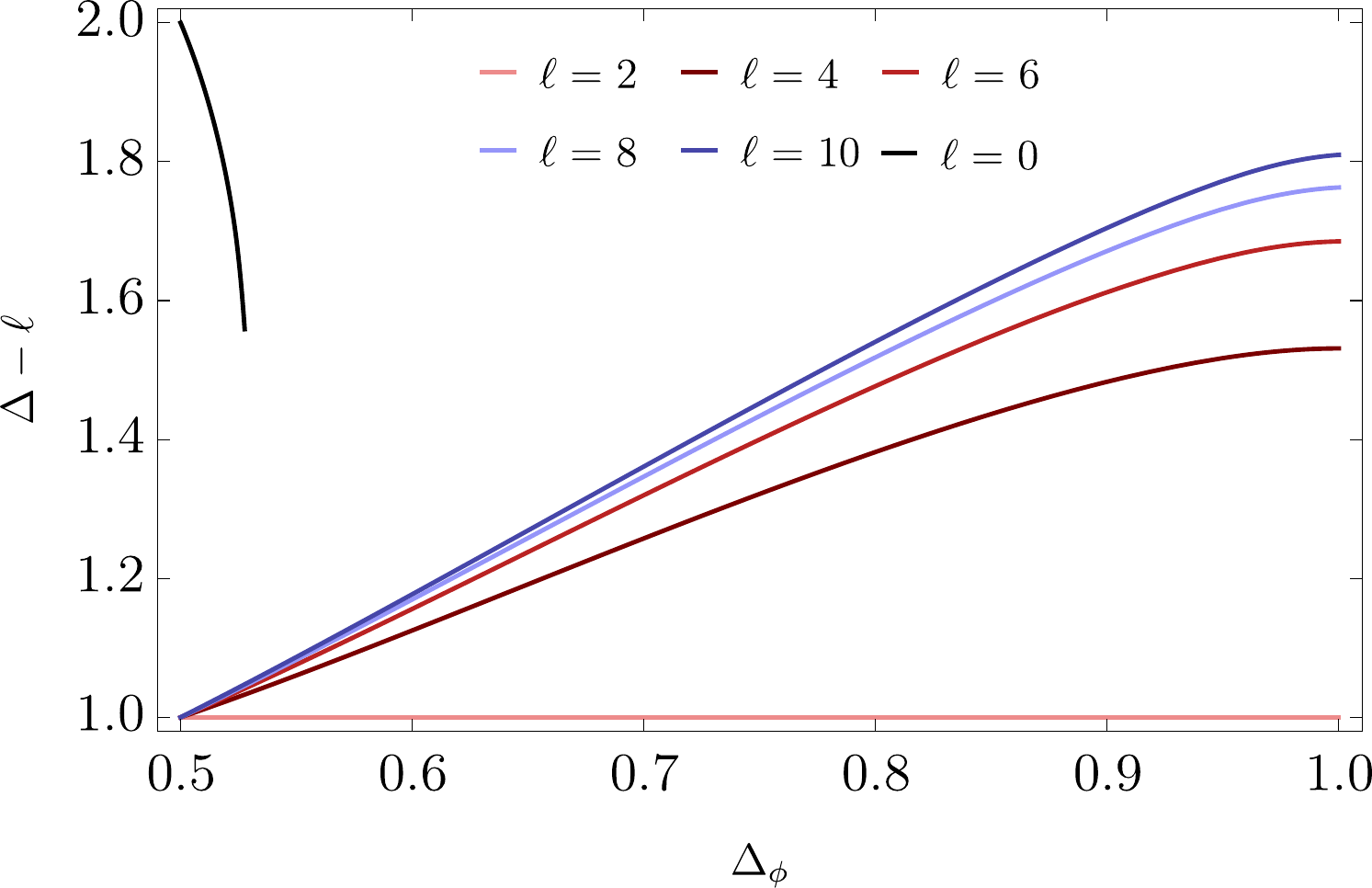}
    \caption{The leading twist operator in the spectrum. $\Delta_{\phi} \in (\frac{1}{2},\frac{3}{4})$ branch is plotted in \cite{CC2022}, here we extrapolate it to $\Delta_{\phi} \in (\frac{1}{2},1)$. When $\Delta_{\phi} = \frac{1}{2}$, the leading twist operator corresponds to the higher spin currents in the critical $O(N)$ model. When $\Delta_{\phi} = 1$, the leading twist operator is identical as in the $q = 3$ bosonic SYK model.}
\label{fig:leading twist}
\end{figure}

By using the ladder kernel function Eq.(\ref{ladder_kernel_function}), one can analyze the operator and the OPE coefficients, Regge intercept (the chaos exponent), and central charge in the spectrum. The operator with the leading twist is plotted in Fig.(\ref{fig:leading twist}).

\begin{widetext}

    \begin{figure}[h]
    \centering
\includegraphics[width = 1\textwidth]{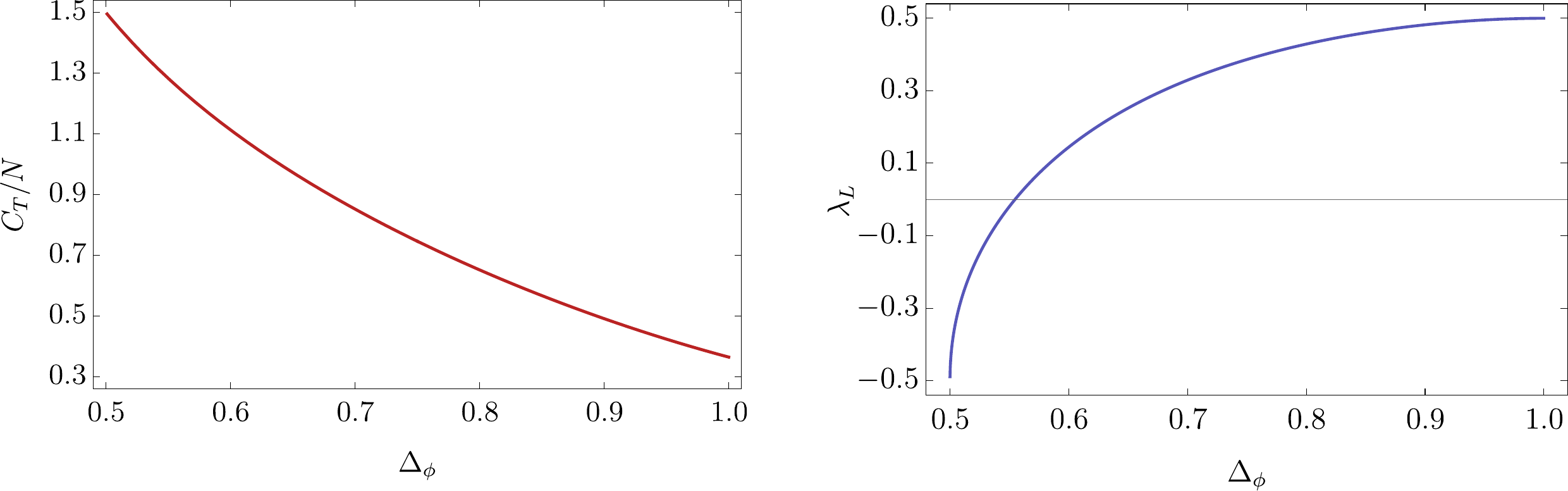}
    \caption{The central charge and the hyperbolic chaos exponent as a function of $\Delta_{\phi}$ in the second branch.}
    \label{fig:BSYKDATA}
\end{figure}
\end{widetext}

It would be interesting to analyze the effects caused by the bulk $\sigma$ field in the boundary spectrum. From the perspective of the bulk, the $\sigma$ field is a free field and does not receive renormalization from the boundary interaction. Hence in the bulk, we have a higher spin symmetry (HS), which means that we are able to find a series of conversed currents $J$ with higher spin in the bulk spectrum, and the scaling dimension of such current would be $\ell+d-2$ in $d$-dimensional bulk without anomalous dimension. In such free theory without the boundary, the higher spin currents correspond to the single-particle states in bosonic Vasiliev theory in $\text{AdS}_{d+1}$.
Due to the presence of the interaction in the boundary, such higher spin currents are not conserved anymore in the bulk perpendicular to the boundary, i.e.
\begin{equation}\label{eom}
    \partial_\nu J^{\nu\mu_1,\cdots,\mu_{\ell}x_\perp} = D_{}^{\mu_1,\cdots, \mu_{\ell}}(\boldsymbol{x})\delta(x_\perp)
\end{equation}
\begin{widetext}

  \begin{figure}
      \centering
      \includegraphics[width = 1\textwidth]{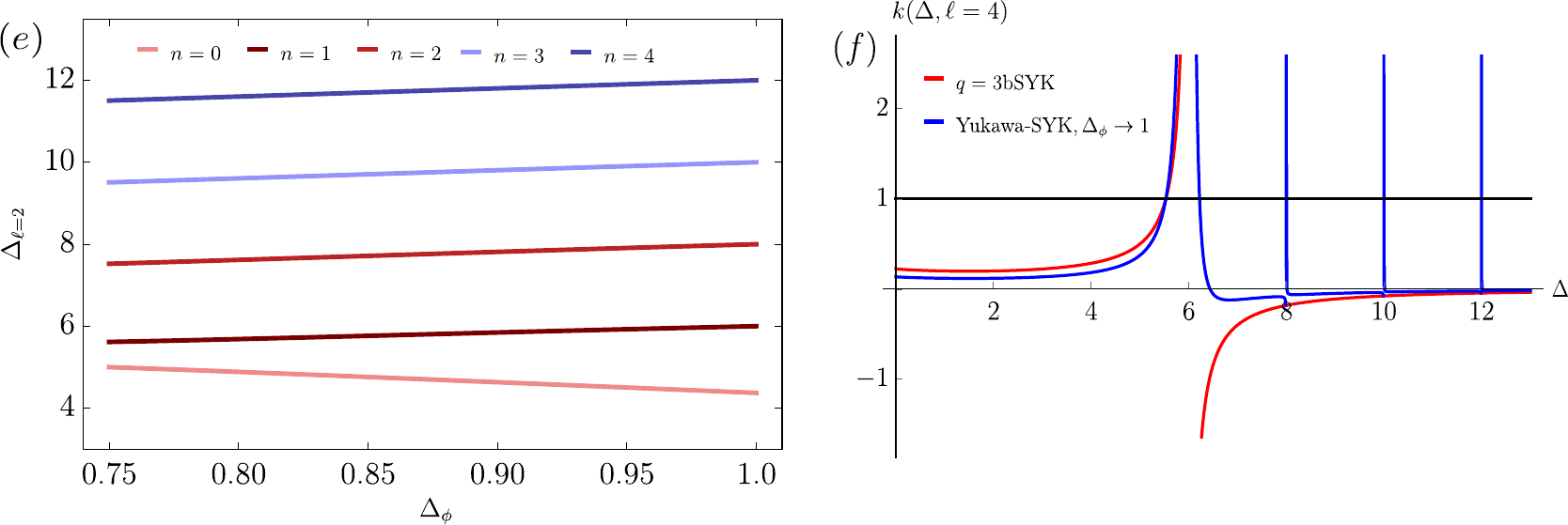}
      \caption{Fig.(e) is the higher twist operator in spectrum $\ell = 2$ as function of $\Delta_{\phi}$ in the second branch. When $\Delta_{\phi} = 1$, we find $\Delta = \ell+2+2n, n\in \mathbb{Z}_{n>0}$, which corresponds to the higher spin displacement operator induced from the bulk CFT. Fig.(f) is the ladder kernel function for $\ell = 4$. The leading twist corresponds to the operator in $q = 3$ bSYK. The higher twist intersections are the displacement operators induced by the bulk. The displacement operator with $n = 0$ has an anomalous dimension.}
      \label{fig:combo5}
  \end{figure}  
\end{widetext}

here we denote the $\boldsymbol{x}$ to be the coordinates in the boundary and $x_\perp$ to be coordinates perpendicular to the boundary and we name the D operator to be the displacement operator. From the perspective of boundary, the spin of the displacement operator ranges from $0$ to $\ell$, which depends on if the tensor index $\mu_i$ belongs to $\boldsymbol{x}$ or $x_\perp$. The divergence equation Eq.(\ref{eom}) further tells that the scaling dimension of the displacement operators and the higher spin currents are the same. Hence we have a tower of displacement operators living in the boundary corresponding to the same higher spin current in the bulk:
\begin{equation}
\begin{aligned}
     D^{\nu_1,\cdots, \nu_{\ell-2}}: (\ell+d-1,s)& , s\leq \ell-2 \quad \text{in boundary} \\
   & \Updownarrow \\
   J^{\mu_1,\cdots \mu_{\ell}}: & (\ell+d-1,\ell)\quad  \text{in bulk}
   \end{aligned}
\end{equation}
The presence of the displacement operators is robust and is protected by the HS symmetry in the bulk. Hence, in a bosonic BCFT with a bulk-free theory one is supposed to find a series of operators with $(\Delta,\ell) = (\ell+2n+d-1,\ell), \ell\in 2\mathbb{Z}_{\geq 0}, n\in \mathbb{Z}_{\geq 0}$ in the $\sigma\sigma\to \sigma\sigma$ channel. The existence of such operators can also be regarded as evidence of the BCFT. In our model, we find such spinning operators in the spectrum at $\Delta_{\phi} = 1$ when for higher twist $n\geq 1$, see Fig.(\ref{fig:combo5}). The operator still has an anomalous dimension for $n = 0$.
The central charge is given by the inverse of the OPE in the stress tensor pole:
\begin{equation}
    |c_{\phi\phi T}|^2 = -\frac{1}{N_1}\underset{\Delta = 3}{\text{Res}}\left(\frac{1-k_{\sigma}(\Delta,\ell)}{1-k(\Delta,\ell)}\frac{\langle\Phi_{\Delta,\ell},F_0\rangle}{\langle\Phi_{\Delta,\ell},\Phi_{\Delta,\ell}\rangle}\right)
\end{equation}

and the central charge is given by:
\begin{equation}
    C_T =  \frac{9\Delta_{\phi}^2}{4|c_{\phi\phi T }|^2} = \frac{9\Delta_{\sigma}^2}{4|c_{\sigma\sigma T}|^2} 
\end{equation}

and is plotted in Fig.(\ref{fig:BSYKDATA}).
When $\Delta_{\phi} = 1$, we find $C_T = \frac{18}{5\pi^2} = \frac{3}{2}C^{q = 3\text{bSYK}}_T$. 

It is also direct to evaluate the Regge intercept, or the hyperbolic chaos exponent $\lambda_L = \ell-1$, by solving
\begin{equation}
    k\left(\frac{3}{2},\ell\right) = 1
\end{equation}
we find that $\lambda_L = 0.5$ at $\Delta_{\phi} = 1$, the same as the $q = 3$ bSYK model, even though the ladder kernel functions Eq.(\ref{ladder_kernel_function})(\ref{ladder_kernel_bsyk}) is different.

\section*{Discussion}
 In this paper, we investigate a series of solvable disordered models with long-range interaction. We analyze the spectrum in the large N and low energy limit. We find a strongly coupled conformal fixed point. The long-range fixed point exhibits many features of long-range CFTs. We also investigate the chaos of the model. The model is a sub-maximal chaotic system compared with strongly-coupled gauge theory, or 0+1d SYK model, and lacks of a gravity dual even in the lowest dimension since the long-range interaction breaks the reparametrization symmetry. In 1+1d, The model can still be a fast scrambler in the light cone limit at small $\alpha$. There exists a critical value of $\alpha$ after which the pole can never contribute and the system can never be fast scrambling. Thus there exists a “scrambling” critical point for the long-range model. The long-range interaction further slows down the butterfly velocity compared with the unitary $1+1$d Lorentzian critical point. In conclusion, this long-range model serves as an analytical platform to investigate the long-range CFTs and the chaos with long-range nature \cite{2020Zhou,2023Zhou,2023Defenu,2023aron,2021Kuwahara,2023Wanisch}
 .
 Motivated by the thought that the correlation of disordered coupling can be regarded as the correlation function of a field from a higher dimension, we proposed a Yukawa SYK-like model with the $\sigma$ field living in the higher dimensional bulk. We find evidence that this model can be reached by tuning the flavor ratio of $\phi$ and $\sigma$. When $\Delta_\phi\to 1$, the spectrum includes the information of the $q = 3$ bosonic SYK model. The chaos exponent and the central charge of the model are the same compared with the bosonic SYK model in three dimensions.  The spectrum of the boundary model consists of displacement operators protected by the higher spin symmetry in the bulk.  

 We provide some thoughts about the models for future study:

 \begin{enumerate}
 
 \item \textit{AdS/BCFT}. As mentioned above, we find a series of operators induced by higher spin currents in the bulk when $\Delta_{\phi}\to 1$.  It would be interesting and accessible to find its holographic dual since the bulk theory is dual to the type-A Vasiliev theory and the boundary theory is an SYK-like model. Does it mean that we are able to find some models dual to the Vasiliev theory in the bulk and JT gravity in the EOW brane?
 
     \item \textit{Late-time chaos}. In section IV, we investigated the chaos behavior when $t$ is smaller than the scrambling time. A natural question is how to describe late-time behavior in sub-maximal chaotic systems \cite{2023Choi,2022Stanford,2022Gu,2023Gao}. The long-range disordered model serves as a concrete example of sub-maximal chaos. Take 0+1d long-range fermionic model for example \cite{2016PhRvDMaldacena}, the effective quadratic action for reparametrization would be
     \begin{equation}
         \sum_{n}\left[\frac{1}{J\beta}n^2(n^2-1)+\alpha(n^2-1)|n|\right]|\epsilon_n|^2
     \end{equation}
     The first term comes from the Schwarzian action, and the second term from the long-range interaction. The long-range term also emerges in the SYK chain \cite{2017Gu}. By formulating in the Kledysh contour, it is direct to get the eikonal action for scramblons \cite{2023Choi}. 
     \item \textit{Non-Fermi liquid}. 
     The SYK and Yukawa SYK model has been introduced in strongly correlated electron systems as one of the toy models in the research of non-Fermi liquid \cite{2022Chowdury,2022Tikhanovskaya,2022Patel,2021Pan,2022Choi,2021keselman}. It is interesting to investigate the possible long-range extension of the SYK-like nFL theory. The pairing vertex of the long-range model in frequency space processes a power law form in a certain limit
     \begin{equation}
         \Delta(i\omega) = \frac{1}{|\omega|^{\gamma}}
     \end{equation}
     which is similar as the $\gamma$ model proposed by Chubukov et al \cite{2020Abanov,2020Wu,2021Wu,Zhang2023,Zhang2023L,2022Zhang}.  
     \item \textit{Measurement induced phase transition}. A novel entanglement transition in the presence of measurement has been discovered in recent years \cite{2019Skinner,2020Choi,2021Jian}. In the model with long-range interaction, the measurement rate has an intriguing interplay with the long-range interaction and hence decides the entanglement structure together \cite{2022Minato,2022Muller,2022Block,2021Sharma}. The long-range disordered model provides a platform for surveying the measurement-induced phase transition from both analytical and numerical perspectives.  
     \item{\textit{Lattice model.}} Lattice models with SYK-like long-range interaction or lattice models with SYK-like interaction localized in some submanifolds. The information scrambling or entanglement entropy would be interesting.
 \end{enumerate}

\section*{Acknowledgement}
XS thanks Chi-Ming Chang, Yingfei Gu, Shao-Kai Jian, Cheng Peng, Zhenbin Yang for helpful discussion and Erez Berg for the internship at Weizmann Institute of Science during the preparation of the manuscript.

\bibliography{apssamp}% Produces the bibliography via BibTeX.

\end{document}